# An Interface for Variational Quantum Eigensolver based Energy (VQE-E) and Force (VQE-F) Calculator to Atomic Simulation Environment (ASE)

M.R. Nirmal, Shampa Sarkar, Manoj Nambiar
TCS Research, Tata Consultancy Services Limited
mr.nirmal@tcs.com, shampa.sarkar@tcs.com, m.nambiar@tcs.com

**Abstract**

The development of quantum algorithms to solve quantum chemistry problems has offered a promising new paradigm of performing computer simulations at the scale of atoms and molecules. Although majority of the research so far has focused on designing quantum algorithms to compute ground and excited state energies and forces, it is useful to run different simulation tasks, such as geometry optimization, with these algorithms as subroutines. Towards this end, we have created an interface for the Variational Quantum Eigensolver based molecular Energy (VQE-E) and molecular Force (VQE-F) code to the Atomic Simulation Environment (ASE). We demonstrate the working of this hybrid quantum-classical interface by optimizing the geometry of water molecule using a native optimizer implemented in ASE. Furthermore, this interface enables one to compare, combine and use quantum algorithms in conjunction with related classical methods quite easily with minimal coding effort.

## Introduction

At a fundamental level, all materials (molecules) are systems of electrons and nuclei with a geometric/structural configuration. Molecules have specific properties based on their possible configurations (states) they can assume, for e.g., 'ground-state' energy, 'excited-state' energy, etc. These properties give molecules characteristic features useful for specific applications, and pave way to synthesize or discover new molecules or material. Calculating these properties at desired precision is not tractable using classical computing methods. Using a quantum computer/algorithm makes it feasible by mapping molecular particle states to qubits (encoding). However, present Noisy Intermediate Scale Quantum (NISQ) devices are limited in qubit counts and qubit coherence (determines length of the program) to run larger molecular simulations, opening opportunities towards hybrid quantum-classical computational algorithms and workflow. In this article, we describe our effort to facilitate the hybrid simulation of molecular systems enabled through integrating the variational quantum Eigensolver (VQE) algorithm based ground state energy calculator to the popular Python framework of atomic simulation environment (ASE).

## Hybrid Quantum Energy and Force Calculator

We use the Variational Quantum Eigensolver (VQE) hybrid algorithm for the computation of ground state energy of a collection of atoms for a specific arrangement in space. This algorithm, proposed by Peruzzo and coworkers (1) and experimentally realized on different quantum hardware platforms (1) (2) (3), works based on the variational principle of quantum mechanics. The trademark of VQE comes from leveraging the capability of quantum computers to prepare the electronic ground state with exponentially fewer resources compared to classical computers. The cost function of VQE is the expectation value of the Hamiltonian describing the system of interest for the generated trial state on quantum hardware. The outcomes of measurement of the trial state are then passed to a classical optimization routine along with the current trial state parameters, which adjusts the parameters according to an updation rule and feeds back the updated set of parameters to the quantum processor to prepare a new state. This circumvents several constraints imposed by NISQ devices in the form of the low coherence times for the qubits, shallow circuits, low-gate fidelity, etc. These state preparation and measurement routines are repeated several times until the energy (expectation value of the Hamiltonian) converges to a minimum.

The forces acting on the atoms can be expressed as the first-order derivative of the ground state energy with respect to their positions. Therefore, we used a finite-difference based numerical differentiation technique to compute the atomic forces where the energy is obtained from the VQE algorithm. The error associated with this numerical estimation of derivatives depends on the type of approximation and the step-size chosen to perturb the atomic arrangement. We chose the central difference approximation to estimate atomic forces since it produces an error proportional to $\Delta x^2$ compared to forward and backward approximations where the error scales as $O(\Delta x)$, where $\Delta x$ is the perturbation of the atoms around the geometry of interest. Other possible ways to compute forces include the quantum algorithms based on Hellman-Feynman theorem (4)(derivative of energy is determined by the expectation value of the derivative of Hamiltonian) to obtain analytical expressions, which could offer more accurate force vectors (5) (6).

We have used the open-source quantum computing toolkit Qiskit (7), provided by IBM, for implementing the VQE algorithm and the statevector simulator to run noiseless quantum simulations. Qiskit is written in Python programming language and is easily compatible with Atomic Simulation Environment (ASE), which is also a Python package. The implementation requires certain set of parameters to be specified for defining the molecular Hamiltonian, choosing an appropriate Active Space (AS), mapping the problem to qubits, creating the trial state ansatz, setting up the classical optimizer. This includes the molecular orbital basis set, number of active electrons and orbitals, various standard fermion to qubit mappings, such as Jordan-Wigner (8), Parity (9) and Bravyi-Kitaev (10) encoding techniques, different hardware-efficient and chemically-inspired ansatz, among others.

**Atomic Simulation Environment (ASE)**

ASE (11) (12) is an open-source suite of Python modules intended to carry out a variety of molecular simulation tasks ranging from geometry optimization, molecular dynamics (MD) to reaction path tracing using different electronic structure calculation methods. ASE contains Python interfaces, which are called Calculators, to a multitude of computational chemistry codes, few of which are listed in Table 1. Some of these codes require heavy numerical computations, and thus are implemented in FORTRAN/C/C++ languages and are not part of ASE, but ASE has Python wrappers interfacing with the actual codes. In contrast, a few calculators are purely implemented in Python and are included within ASE package.

Table 1: List of different calculator interfaces available in ASE. They are grouped based on their types: Density Functional Theory (DFT), Linearized Augmented Planewave (LAPW), Hartree Fock (HF), Effective Medium Theory (EMT), Classical Molecular Dynamics (CMD)

| Code | Description | Type |
| --- | --- | --- |
| ABINIT | Planewave pseudopotential code | DFT |
| DFTB+ | DFT based tight binding code | DFT |
| FLEUR | Full potential LAPW code | DFT, LAPW |
| EMT | Effective Medium Theory potential (written in Python) | EMT |
| Asap | Highly efficient EMT code (written in C++) | EMT |
| GPAW | Grid-based real-space PAW code | DFT, HF |
| LAMMPS | Classical Molecular Dynamics code | CMD |
| Siesta | Linear combination of atomic orbitals pseudo potential code | DFT |

At the core of ASE framework lies the Atoms class. One can define a molecule, ion, a lattice or any collection of atoms using the Atoms class and store all the related properties including atomic numbers, positions, masses, charges, magnetic moments, and velocities. An ASE calculator interface is then attached to the Atoms object that enables the straightforward calculation of potential energies and forces of the atomic arrangement according to the underlying algorithm of the calculator. One can leverage the dedicated modules within ASE to further perform more complex simulation tasks, such as molecular dynamics simulation, by easily calling the associated Python class. The module then derives the energy and forces as evaluated by the assigned calculator of the Atoms object to evolve the atomic configuration. The various simulations supported by ASE currently include molecular dynamics with different controls such as thermostats, geometry optimization using atomic forces and genetic algorithm, transition state optimization using the nudged elastic band method (NEB), global geometry optimization using basin hopping, vibrational mode analysis among many others. In addition, ASE provides a database module using which an user can conveniently store and retrieve the results of the simulations performed in a number of file formats (more than 65 file formats are being supported). The overall workflow in using the hybrid quantum calculator with ASE to carry out different simulation tasks is shown in Figure 1.

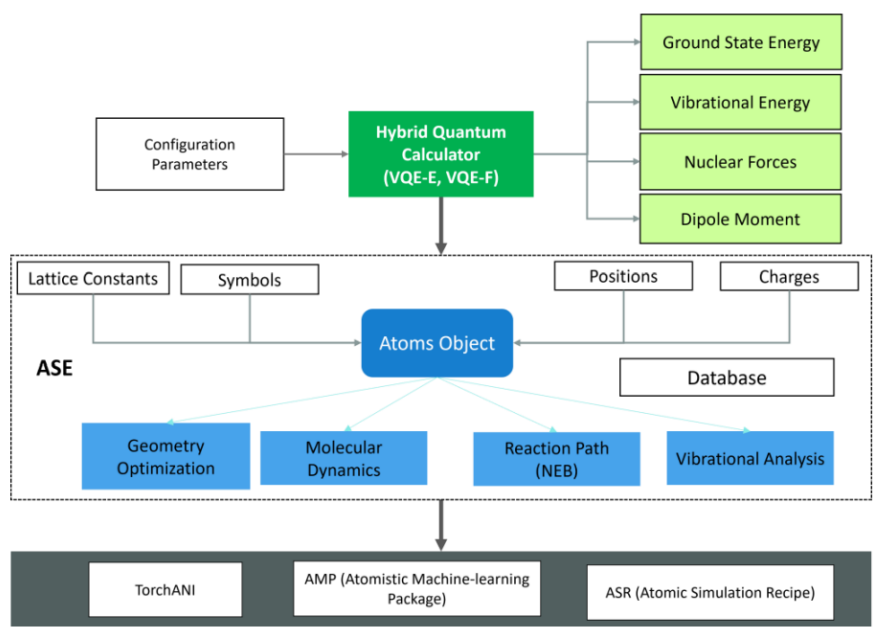

*Figure 1: A schematic illustrating the various modules within ASE and the workflow in using the hybrid VQE-E and VQE-F calculator with ASE*

## Results

For the purpose of demonstrating the functionality of the hybrid quantum interface within the ASE framework, we attempted to optimize the structure of water ($H_2O$) molecule using the in-built local optimizers available in ASE. The steps undertaken for this simulation task are:

i. Instantiate the ASE Atoms class to represent a water molecule with initial geometry as given by classical HF level of theory.
ii. Import the hybrid quantum energy and force calculator and configure it by passing the necessary parameters as a Python dictionary through a simple text file.
iii. Attach the calculator to the Atoms object created in step i.
iv. Import and create an instance of the local optimizer, say BFGS, from the module ase.optimize
v. Pass the Atoms object to the optimizer and run the optimizer until the force on all atoms becomes below a predefined threshold ($10^{-5}\ Hartree/Angstorms$)

To illustrate the flexibility of the interface to compute energy and forces in different molecular orbital (MO) basis sets, we perform the geometry optimization in both 6-31G* and cc-pVTZ MO basis sets. The results are presented in Table 2 and the trajectory of structure optimization with 6-31G* is shown in Figure 2.

Table 2: The optimized geometries and minimum energies as obtained from the ASE interface to hybrid quantum calculator

| MO Basis set | Optimized rOH (Å) | Optimized aHOH (degrees) | Minimum Energy (Hartrees) | Iterations |
|---|---|---|---|---|
| 6-31G* | 0.94767139 | 105.6029291 | -76.009489361 | 7 |
| cc-pVTZ | 0.94065276 | 106.006883 | -76.057836158 | 5 |

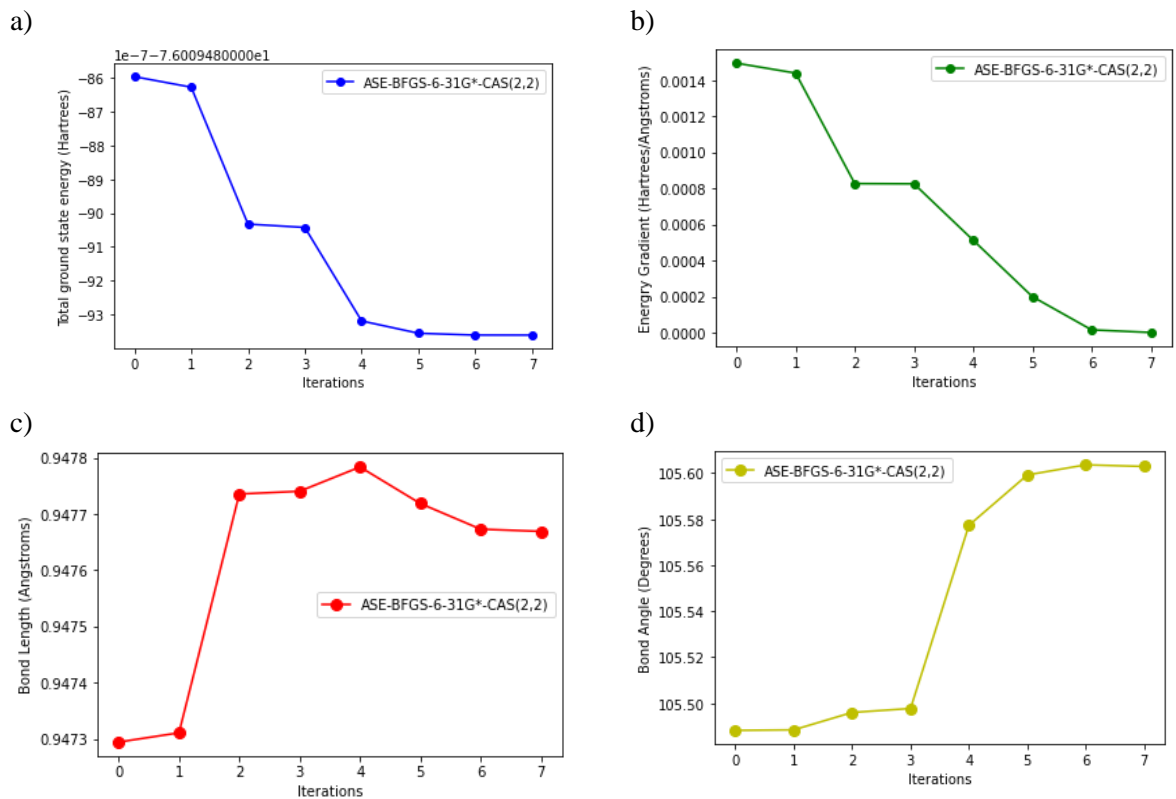

Figure 2: Trajectory of geometry optimization of $H_2O$ in 6-31G* basis using hybrid quantum interface within the ASE framework: a) Total ground state energy vs iterations b) Norm of the force vector vs iterations c) O-H bond length vs iterations d) H-O-H bond angle vs iterations

## Conclusion and Future Work

In this work, we present an initial version of an interface between Atomic Simulation Environment and Variational Quantum Eigensolver based molecular Energy (VQE-E) and molecular Force (VQE-F) code. The functionality of the interface is illustrated by performing a structure optimization procedure on water molecule using a native optimizer of ASE. One can configure the calculator to evaluate other molecular properties of interest such as dipole moment, angular momentum and vibrational energies once efficient quantum algorithms to determine these physical quantities are developed. In recent years, many open-source deep learning and machine learning (DL/ML) packages, such as TorchANI (13) (14), Amp (15), have been developed to work on top of ASE. They use ASE as an interface to communicate with different electronic structure codes to train their DL/ML models and possibly infer more accurate estimates of molecular properties.

**Acknowledgement**

The authors thank Sriram Goverapet Srinivasan for his valuable insights and guidance throughout the research work.